\documentstyle[epsfig]{article}
\title
{\large {\bf
ON THE REACTION $ep \to ep \gamma$
\thanks{Talk at The Vth International Summer School-Seminar {\it Actual
Problems of Partile Physics} (Belarus, Gomel, July 30 - August 8 1999)}
} }
\author {M.V. Galynsky and M.I. Levchuk \thanks{E-mails:
galynski@dragon.bas-net.by, levchuk@dragon.bas-net.by}  \\
{\it Institute of Physics,
Belarusian Academy of Sciences, Minsk}
}
\date{}
\begin{document}
\newcommand{\MEV}{\mbox{MeV}}
\newcommand{\beq}{\begin{equation}}
\newcommand{\eeq}{\end{equation}}
\newcommand{\beqn}{\begin{eqnarray}}
\newcommand{\eeqn}{\end{eqnarray}}
\maketitle

\begin{abstract}
We have studied the reaction $ ep \to ep \gamma$ in the kinematics
corresponding to electron scattering at small angles and photon
scattering at large angles, where proton bremsstrahlung dominates.
The analysis is based on the direct evaluation method of the matrix
elements in the so-called diagonal spin basis. The results of
numerical calculations for electron beam energy $E_{e}=200\, \MEV$ in
the above kinematics show that the relative contribution of the
Bethe-Heitler and interference terms to the reaction cross section is less
than 10 \%, and the cross section for the reaction $ep \to ep \gamma$ is
quite sensitive to the proton polarizability.
Owing to the factorization of the  squared electric and magnetic form
factors of the proton, a compact expression has been obtained for
the differential cross section of the Bethe-Heitler emission of a
linearly polarized photon by an electron, taking into account the
proton recoil and form factors.
A covariant expression has been obtained for the lepton tensor in
which contributions of states with transverse and longitudinal
polarizations of the virtual photon are separated.
\end{abstract}

\section{\bf The reaction $ep \to ep \gamma$ and the proton polarizability}

There has recently been much interest in studying Compton scattering on
nucleons at low and intermediate energies. The motivation is
that the fundamental structure constants of the nucleon, the
electric and magnetic polarizabilities, can be determined in this
process. The nucleon polarizabilities contain important information
about the nucleon structure at large and intermediate distances, in
particular, about the radius of the quark core, the meson cloud, and
so on. A detailed discussion of these questions can be found in
\cite{b58,b59}.  Knowledge of the amplitudes for Compton scattering
on nucleons is also required to interpret the data on photon
scattering off nuclei.  For example, such studies can
answer the question of in what degree the electromagnetic
properties of free and bound nucleons differ.

All the experimental results on the proton polarizabilities have been
obtained from data on elastic $\gamma p$ scattering below pion
photoproduction threshold \cite{macgib}. However, it has recently
been shown that measurements of the proton polarizabilities at the
Novosibirsk storage ring with electron beam energy of $200\, \MEV$
using an internal jet target appear to be very promising. As proposed
in \cite{b60}, this can be done using the reaction
\beq
e^{-}(p_{1})+p^{+}(q_{1}) \to e^{-}(p_{2}) + p^{+}(q_{2}) + \gamma(k)
\label{5.1}
\eeq
in the kinematics corresponding to electron scattering at small angles and
photon scattering at large angles, i.e. in conditions of
small 4-momentum transfer from the initial electron to
the final photon and proton. In the lowest order of perturbation
theory, the process (\ref{5.1}) is described by three graphs
shown in Fig.1.

The first two (a) and (b) correspond to electron bremsstrahlung
(Bethe-Heitler graphs), and the third (c) corresponds to proton
bremsstrahlung (graph with virtual Compton scattering (VCS) on a
proton).  The kinematics described above was chosen for the following
reasons.  First, the subprocess of real Compton scattering (RCS) on
the proton is realized in it since at small electron scattering
angles the virtual photon with 4-momentum $r = p_{1} - p_{2}$ (see
Fig.1) becomes almost real. Here the quantity $|r| = \sqrt{-(p_{1} -
p_{2})^{2}}$ turns out to be small, $|r| \sim m$, where $m$ is the
electron mass. Second, for electron scattering at small angles and
photon scattering at large angles, the contribution of the
graph corresponding to proton bremsstrahlung dominates, being
several orders of magnitude larger than the contribution of the
Bethe-Heitler graphs to the cross section for the process (\ref{5.1})
\cite{b61}. This is the main requirement needed to separate the
subprocess of Compton scattering on the proton \cite{b60} in the
reaction $e p\to ep \gamma$.

\vspace{1.7cm}
\begin{figure}[ht]
\centerline{\epsfbox[10 10 550 700]{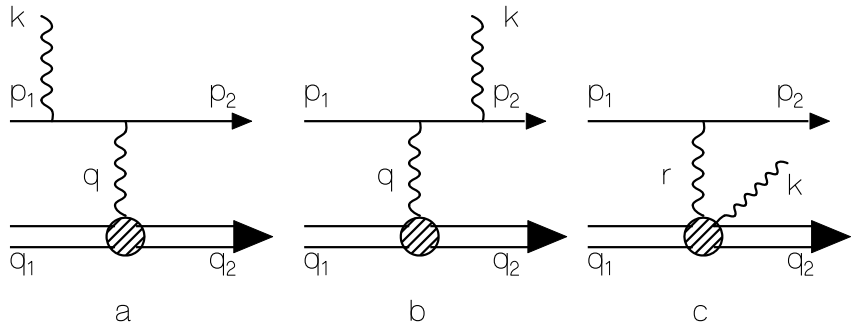}}
\vspace{-22cm}
\caption{ Graphs corresponding to the reaction $ep \to ep\gamma$. }
\end{figure}

The estimates in the framework of the method of equivalent photons
for a scalar model \cite{b60} showed that the reaction (\ref{5.1})
offers a good possibility of obtaining high-statistics data on the
Compton scattering cross section and the proton polarizability.
Measurement of the electric ($\alpha_{p}$) and magnetic
($\beta_{p}$) polarizabilities of the proton  with higher accuracy
than in earlier studies is one of the most important problems to be
solved by experiments in the near future \cite{b62,b63}.

However, to obtain high-statistics data on the cross section for $\gamma
p$ scattering and the proton polarizability it is essential to use a
theoretical model more accurate than that of \cite{b60}. It must
include both the spin properties of the particles and
parameters characterizing  the electromagnetic structure
of the hadron. The model can be based on the result of \cite{b64},
where a general calculation of the reaction $e p\to e p \gamma$ was
performed. The cross section was expressed in terms of 12 form
factors corresponding to the VCS subprocess on the proton (i.e., the
contribution of the graph in Fig.1c) and two form factors
corresponding to the Bethe-Heitler graphs.

The differential cross section for the reaction $e p\to e p \gamma $ in the
above kinematics was calculated in \cite{b65}. It was expressed in terms of
the six invariant amplitudes for RCS \cite{b58,b66}, and also the electric
and magnetic form factors of the proton \cite{b10}.

The  matrix element corresponding to the sum of the two Bethe-Heitler
graphs (a) and (b) in Fig.1 reads
\beq
M_{1} = \overline{u}(p_{2}) Q^{\mu}_{e} u(p_{1}) \cdot \overline{u}(q_{2})
\Gamma_{\mu}(q^{2}) u(q_{1}) \; {1 \over q^{2}} \; ,
\label{5.2}
\eeq
\beq
Q^{\mu}_{e} = \gamma^{\mu} \; {\hat p_{1} - \hat k + m \over - 2 p_{1} k }
\; \hat e + \hat e \; {\hat p_{2} + \hat k + m \over  2 p_{2} k } \;
\gamma^{\mu} \; ,
\label  {5.3}
\eeq
\beq
\Gamma_{\mu}(q^{2}) = f_{1} \; \gamma_{\mu} + {\mu_{p} \over 4M} \; f_{2} \;
( \; \hat q \gamma_{\mu} - \gamma_{\mu} \hat q \; ) \; ,
\label {5.4}
\eeq
where $u(p_{i})$ and $u(q_{i})$ are the bispinors of electrons and protons
with 4-momenta $p_{i}$ and $q_{i}$, $p_{i}^{2} = m^{2}, \; q_{i}^{2}
= M^{2}, \; \overline{u}(p_{i})\; u(p_{i})=2m, \; \overline{u}(q_{i}) \;
u(q_{i})= 2M, \; (i = 1,2), \; \hat{k}=k_{\mu} \gamma^{\mu}, \,
\gamma^{\mu}$ are the Dirac matrices, $\gamma^5=-i \gamma^0 \gamma^1
\gamma^2 \gamma^3, \, \gamma^{5+}=\gamma^5; \;\mu_{p}, \;  f_{1}$, and $f_{2}$ are
respectively the anomalous magnetic moment and the Dirac and Pauli form
factors of the proton \cite{b10}, $q = q_{2}-q_{1}$ is the momentum transfer,
$e$ is the polarization 4-vector of a photon with momentum $k, \; e k=
k^{2}=0$, and $M$ is the proton mass.

In the limit of interest $ |r| \sim m$, the matrix element corresponding
to the graph of Fig.1c is expressed in terms of the six invariant RCS
amplitudes $T_{i} \; (i = 1,2, \ldots, 6) $. It has the form \cite{b64}
\beq
M_{2} = \overline{u}(p_{2}) \gamma^{\mu} u(p_{1}) \cdot \overline{u}(q_{2})
M_{\mu \nu} e^{\nu} u(q_{1}) \; {1 \over r^{2}} \; ,
\label {5.5}
\eeq
\beq
M_{\mu \nu} = { C_{\mu} C_{\nu} \over C^{2}} \; ( T_{1} + T_{2} \hat K ) +
{ D{\mu} D{\nu} \over D^{2}} \; ( T_{3} + T_{4} \hat K ) +
\nonumber
\eeq
\beq
+ { (C_{\mu} D_{\nu} - C_{\nu} D_{\mu}) \over D^{2}} \; \gamma^{5} \;
T_{5} \;  + \; { (C_{\mu} D_{\nu} + C_{\nu} D_{\mu}) \over D^{2}} \;
T_{6} \hat D \; .
\label {5.6}
\eeq
The tensor $M_{\mu \nu}$ is constructed using a set of four mutually
orthogonal 4-vectors $C, D , B$, and $K$:
$$
K = 1/2 \; ( r + k ) \; , \; Q = 1/2 ( r - k ) \; , \; R = 1/2 ( q_{1} + q_{2} ) \; ,
$$
\beq
C = R - { ( R K )\over K^{2}} \; K - { ( R B )\over B^{2}} \; B \; , \;
B = Q - { ( Q K )\over K^{2}} \; K  \; ,
\label {5.7}
\eeq
$$
D_{\mu} = \varepsilon_{\mu \nu \rho \sigma} K^{\nu} B^{\rho} C^{\sigma} \; ,
$$
and it satisfies the requirements of parity conservation and gauge
invariance:
\beq
M_{\mu \nu} k^{\nu} = r^{\mu} M_{\mu \nu} = 0 \; .
\label {5.8}
\eeq

In the unpolarized case it is most efficient to use the standard approach
\cite{b10} for calculation of the differential cross section of
the process (\ref{5.1}) together with evaluation of matrix
elements in the diagonal spin basis (DSB) \cite{b36}-\cite{b47}. In the
DSB, the spin 4-vectors $s_{1}$ and $s_{2}$ of particles with
4-momenta $p_{1}$ and $p_{2}$ ($s_{1} p_{1} = s_{2} p_{2} = 0 , s_{1}
^{2} = s_{2} ^{2} = - 1 $) belong to the hyperplane formed by the
4-vectors $p_{1}$ and $p_{2}$:
\beq
s_{1} = - \; { (v_{1} v_{2}) v_{1} - v_{2} \over \sqrt{ ( v_{1}v_{2}
)^{2} - 1 }} \; \; , \; \; s_{2} =  { ( v_{1} v_{2}) v_{2} - v_{1}
\over \sqrt{ ( v_{1}v_{2} )^{2} - 1 }} \; \; ,
\label {4}
\eeq
where $v_{1} = p_{1}/m_1$ and $ v_{2} = p_{2}/m_2 $. To find the probability
for the process (\ref{5.1}) it is sufficient to calculate the matrix
elements of the electron and proton currents
\beq
( J^{\pm\delta,\delta}_{e} )_{\mu} = \overline{u}^{\pm\delta}(p_{2})
\gamma_{\mu} u(p_{1})^{\delta}  \; ,
\label {5.15}
\eeq
\beq
( J^{\pm\delta ',\delta '}_{p} )_{\mu} = \overline{u}^{\pm \delta '}(q_{2})
\Gamma_{\mu}(q^{2}) u^{\delta '}(q_{1}) \; ,
\label {5.16}
\eeq
and also the quantity
\beq
X^{\pm\delta ',\delta '}_{\mu} = \overline{u}^{\pm \delta '}(q_{2})
M_{\mu \nu} e^{\nu} u^{\delta '}(q_{1})  \; .
\label {5.17}
\eeq
The calculations give \cite{b36}-\cite{b47}:
\beq
( J^{\delta,\delta}_{e} )_{\mu} = 2 m ( a_{0} )_{\mu} \; , \;
( J^{-\delta,\delta}_{e} )_{\mu} = - 2 \delta y_{-} ( a_{\delta} )_{\mu} \; ,
\; y_{-}=\sqrt{-p_{-}^{2}} /2 \; ,
\label {5.18}
\eeq
\beq
( J^{\delta ',\delta '}_{p} )_{\mu} = 2 g_{e} M ( b_{0} )_{\mu} \; , \;
( J^{-\delta ',\delta '}_{p} )_{\mu} = - 2 \delta^{'} y_{-}^{'} g_{m} (
b_{\delta '} )_{\mu} \; ,\; y_{-}^{'}=\sqrt{-q_{-}^{2}}/2,
\label {5.19}
\eeq
where
\beq
a_{0} = p_{+}/\sqrt{p_{+}^{2}} \; , \; a_{3} = p_{-}/\sqrt{-p_{-}^{2}} \; ,
\; a_{2} = [a_{0} \cdot a_{3}]^{\times} k /\rho \; , \;  a_{1} = [a_{0}
\cdot a_{3}]^{\times} a_{2} \; ,
\label {5.9}
\eeq
\beq
p_{\pm} = p_{2} \pm p_{1} \; , \; a_{\pm \delta} = a_{1} \pm i \delta a_{2} \; ,
 \; \delta = \pm 1 \; , \; a_{2} k = 0 \; , \; a_{1}^{2} =
a_{2}^{2} = a_{3}^{2} = - a_{0}^{2} = - 1 \; .
\label{5.9a}
\eeq
\beq
b_{0} = q_{+}/\sqrt{q_{+}^{2}} \; , \; b_{3} = q_{-}/\sqrt{-q_{-}^{2}} \; ,
\; b_{2} = [b_{0} \cdot b_{3}]^{\times} k /\rho^{'} \; , \;  b_{1} = [b_{0}
\cdot b_{3}]^{\times} b_{2} \; ,
\label {5.10}
\eeq
\beq
q_{\pm} = q_{2} \pm q_{1} \; , \; b_{\pm \delta^{'}} = b_{1} \pm i
\delta^{'} b_{2} \; ,  \; \delta^{'} = \pm 1 \; , \; b_{2} k = 0 \; ,
\; b_{1}^{2} = b_{2}^{2} = b_{3}^{2} = - b_{0}^{2} = - 1 \; .
\label{5.11}
\eeq
In Eqs.(\ref{5.9}), (\ref{5.10}) and below a dot between any two 4-vectors
$a$ and $b$, square parentheses and symbol "$^{\times}$" stands for dyadic
product of vectors (but not scalar product) $a \cdot b=(a \cdot b)_
{\mu\nu} =(a)_{\mu}(b)_{\nu}$, alternating dyadic $[a \cdot b]=a \cdot b
-b \cdot a$ and dual operation $[a \cdot b]^{\times}=([a \cdot b]^
{\times})_{\mu\nu}=1/2\varepsilon_{\mu\nu\rho\sigma}([a \cdot b])^
{\rho\sigma}=\varepsilon_{\mu\nu\rho\sigma}(a)^{\rho}(b)^{\sigma}$,
respectively, $\varepsilon_{\mu\nu\rho\sigma}$ is the Levi-Civita symbol
($\varepsilon_{0123}=-1$); $\rho$ and $\rho^{'}$ are determined from the
normalization conditions (\ref{5.9a}) and (\ref{5.11}), finaly, $g_{e}$
and $g_{m}$ are just electric and magnetic form factors of the proton
(Sachs form factors):
\beq
g_{e} = f_{1} + \mu_{p} {q^{2}\over 4 M^{2}} \; f_{2} \; , \; g_{m} =
f_{1} + \mu_{p} \; f_{2} \; .
\label {5.20}
\eeq
Therefore, in the DSB the matrix elements of the proton current for
spin-non-flip and spin-flip transitions are expressed in terms of
the electric  $g_{e}$ and magnetic $g_{m}$ form factor,
respectively (see \cite{b11}).

Once the matrix elements of the proton current (\ref{5.16}) have
been determined, the calculation of the contribution of the two
Bethe-Heitler graphs reduces to the calculation of VCS on the
electron \cite{b65,b40a,b47}:
\beq
|M_{1}^{\pm\delta ',\delta '}|^{2} = {1\over q^{4}} |\overline{u}(p_{2})
\left (\hat J_{p}^{\pm\delta ',\delta '} {\hat p_{1} - \hat k + m \over
- 2 p_{1} k } \hat e + \hat e \; {\hat p_{2} + \hat k + m \over  2 p_{2} k }
\hat J_{p}^{\pm\delta ',\delta '} \right ) u(p_{1}) |^{2} \; .
\label {5.21}
\eeq
Denoting the result of averaging and summing the expression
$|M_{1}^{\pm\delta ',\delta '}|^{2}$ over the polarizations of the initial
and final particles by $Y_{ee}$, one obtains \cite{b65,b40a,b47}:
\beq
Y_{ee} = 1/4 \; \sum_{\delta ' e} \; Tr\{ \; ( \hat p_{2} + m ) \;
\widehat Q_{e}^{ \pm\delta ',\delta '} \; ( \hat p_{1} + m ) \;
\widehat{\overline{Q}}_{e}^{\pm\delta ',\delta '} \; \} / q^{4} \; ,
\label {5.22}
\eeq
where $ \widehat Q_{e}^{ \pm\delta ',\delta '} = ( Q_{e}^{\mu} ) \, (
 J_{p}^{\pm\delta ',\delta '} )_{\mu}$ is the operator in parentheses between
the electron bispinors $ \overline{u}(p_{2})$ and $u(p_{1})$ in Eq.
(\ref{5.21}), and $\widehat{\overline{Q}}_{e}^{\pm\delta ',\delta '}
= \gamma_{0}\, (\widehat Q_{e}^{ \pm\delta ',\delta '} )^{+} \gamma_{0}$.
Owing to the factorization of the electric and magnetic form factors
$g_{e}$ and $g_{m}$ in (\ref{5.19}), the Bethe-Heitler term in the cross
section for the reaction $ep \to ep\gamma \; Y_{ee}$ (\ref{5.22})
contains only the squares of the Sachs form factors
(see \cite{b65,b36,b40a,b47,b67,b68}).

Similarly, the calculation of the contribution from the graph in
Fig.1c reduces to the calculation of quasi-real Compton scattering on
the proton.  Using the expressions for the electron current
(\ref{5.18}), one has
\beq
\mid M_{2}^{\pm\delta,\delta}\mid^{2} = {1\over r^{4}} \mid \overline{u}
(q_{2}) \; \widehat Q_{p}^{ \pm\delta , \delta } \; u(q_{1}) \mid^{2} \; ,
\label {5.23}
\eeq
where $\widehat Q_{p}^{ \pm\delta, \delta }=(J^{\pm\delta,\delta}_{e})^{\mu}
 \;M_{\mu \nu} e^{\nu}$. Denoting the result of averaging and summing Eq.
(\ref{5.23}) over the polarizations of the initial and final particles
by $Y_{pp}$, we obtain \cite{b65}:
\beq
Y_{pp} = 1/4 \; \sum_{\delta e} \; Tr \{ \; ( \hat q_{2} + M ) \;
\widehat Q_{p}^{ \pm\delta , \delta } \; ( \hat q_{1} + M ) \;
\widehat{\overline{Q}}_{p}^{\pm\delta , \delta } \; \} / r^{4} \; ,
\label {5.24}
\eeq
where $ \; \widehat{\overline{Q}}_{p}^{~\pm\delta, \delta }=\gamma^{0}\;
(\widehat Q_{p}^ { \pm\delta , \delta } )^{+} \gamma^{0}$. Finally, to
calculate the interference term in the case of unpolarized particles
\beq
Y_{ep} = 1/4 \; \sum_{\delta,\delta ',e} \; 2 Re \; M_{1} \; M_{2}^{\ast}
\label {5.25}
\eeq
we shall use the matrix elements of the proton current (\ref{5.19}) and
also the 4-vectors $X^{\pm\delta ',\delta '}_{\mu}$ (\ref{5.17}), which
have the form \cite{b65}
$$
X^{-\delta ',\delta '}_{\mu} = -2 \delta^{'} y_{-}^{'} b_{1} k \left (
{ C_{\mu} C_{\nu} \over C^{2}} T_{2} + { D{\mu} D{\nu} \over D^{2}} T_{4}
 + i \delta^{'} y_{+}^{'} y_{-}^{'} { (C_{\mu} D_{\nu} + C_{\nu} D_{\mu})
 \over D^{2}} T_{6} \right ) e^{\nu} \; ,
$$
\beq
X^{\delta ',\delta '}_{\mu} = 2 \left ( y_{+}^{'} \left (
 { C_{\mu} C_{\nu} \over C^{2}} \left ( T_{1} + {\nu_{1} M \over 1 - \tau} T_{2} \right )
 + { D{\mu} D{\nu} \over D^{2}} \left ( T_{3} + {\nu_{1} M \over 1 - \tau} T_{4} \right )
 \right ) \right. +
\label {5.26}
\eeq
$$
+ \left. \delta^{'} y_{-}^{'} { (C_{\mu} D_{\nu} - C_{\nu} D_{\mu})
\over D^{2}} \; T_{5}  \right )  e^{\nu} \; ,
$$
where $y_{+}^{'}=\sqrt{q_{+}^{2}}/2=M \sqrt{1-\tau}\;,\;\tau =q^{2}/4 M^{2}$
and $\nu_{1}=k q_{+}/2M^{2}$. As a result, one has for the matrix
element $M_{2}$ (\ref{5.5})
\beq
M_{2} = \overline{u}(p_{2}) \hat{X}^{\pm \delta ',\delta '} u(p_{1}) / r^{2} \; ,
\label {5.27}
\eeq
and Eq.(\ref{5.25}) reduces to the trace \cite{b65}:
\beq
Y_{ep} = 1/4 \sum_{\delta,\delta ',e} 2 Re \; \{Tr\; ( ( \hat p_{2} + m )
 \widehat Q_{e}^{ \pm\delta ',\delta '}  ( \hat p_{1} + m )
 \hat{\overline{X}}^{\pm\delta ',\delta '} )\} / q^{2} /r^{2} \; ,
\label {5.28}
\eeq
where $\hat{X}^{\pm \delta ',\delta '} = \gamma^{\mu} \; X^{\pm\delta ',
\delta '}_{\mu}$ and $\hat{\overline{X}}^{\pm\delta ',\delta '} =
( X^{\pm\delta ',\delta '}_ {\mu} )^{\ast} \gamma^{\mu} $. The interference
term $Y_{ep}$ (\ref{5.28}) is a linear combination of the proton electric and
magnetic form factors, because the operators $\widehat Q_{e}^{\pm\delta ',
\delta '}$ are expressed linearly in terms of the matrix elements of the
proton current: $\widehat Q_{e}^{ \pm\delta ',\delta '} = (Q_{e})^{\mu}
( J^{\pm\delta ',\delta '}_{p} )_{\mu} \; $, (see Eqs. (\ref{5.3}) and
(\ref{5.19})).
Therefore, the problem of finding the probability for the reaction $ e p
\to e p \gamma$ in this approach has been reduced to calculations of the
traces (\ref{5.22}), (\ref{5.24}), and (\ref{5.28}), which were done
making the use of the program REDUCE. For the differential
cross section we then obtained \cite{b65,b40a}:
\beq d \sigma={\alpha^{3} \mid T \mid ^{2} \delta^{4}(p_{1}+q_{1}-p_{2} -
q_{2} - k ) \over 2 \pi^{2} \sqrt{ (p_{1} q_{1})^{2} - m^{2} M^{2}}}
\; \; { d^{3} \vec p_{2} \over 2 p_{20}} \; {d^{3} \vec q_{2} \over 2
q_{20}} {d^{3} \vec k \over 2 \omega} \; \; ,
\label {5.29}
\eeq
\beq
\mid T \mid^{2} = 1/4 \; \sum_{pol} \mid M_{fi} \mid^{2} = Y_{ee} +
Y_{ep} + Y_{pp} \; , \label {5.30} \eeq \beq Y_{ee} = {8 M^{2}\over
q^{4}} \; ( \; g_{e}^{2} \; Y_{I} + \tau \; g_{m}^{2} \; Y_{II} \; )
\; , \label {5.31}
\eeq
$$ Y_I= -\;\frac{\lambda_1}{\lambda_2}-\frac{\lambda_2}{\lambda_1}-
\frac{m^2q^2}2\left(\frac 1{\lambda_1}-\frac 1{\lambda_2}\right)^2-
\frac {r^2q^2}{2\lambda_1\lambda_2}
$$
\beq
-\; {m^{2} \over 2 M^2(1-\tau)} \left ( {p_{1} q_{+} \over \lambda_{2}} -
{p_{2} q_{+} \over \lambda_{1}} \right )^{2} - {\tau \over (1 - \tau)}
{( (p_{1} q_{+})^{2} + (p_{2} q_{+})^{2} ) \over \lambda_{1} \lambda_{2}} \; ,
\label {5.32}
\eeq
$$
Y_{II}= -\; \frac{\lambda_1}{\lambda_2}-\frac{\lambda_2}{\lambda_1}-
\frac{m^2q^2}2\left(\frac 1{\lambda_1}+\frac 1{\lambda_2}\right)^2-
\frac {r^2q^2}{2\lambda_1\lambda_2}
$$
\beq
+\; {m^{2} \over 2 M^2(1-\tau)} \left ( {p_{1} q_{+} \over \lambda_{2}} -
{p_{2} q_{+} \over \lambda_{1}} \right )^{2} + {\tau \over (1 - \tau)}
{( (p_{1} q_{+})^{2} + (p_{2} q_{+})^{2} ) \over \lambda_{1} \lambda_{2}}
\label {5.33}
\eeq
$$
- \; 2 \left ( {m^2 \over \lambda_{1}} - {m^2 \over \lambda_{2}} \right )^{2} + 4
\; m^2 \left ( {1 \over \lambda_{1}} - {1 \over \lambda_{2}} \right ) \; ,
\nonumber
$$
$$
Y_{ep}=-\; {32 M^3\over r^2 q^2 (4 \nu_{4}^2 - \nu_{2}^2)} \Biggl \{
g_{e} Re \left [ y_{1} \left ( T_{1} + {\nu_{1} M\over 1 - \tau }
 T_{2} \right ) + y_{2} \left ( T_{3} + {\nu_{1} M\over 1 - \tau } T_{4}
 \right ) \right ] \biggr.
$$
\beq
\biggl. + \tau g_{m} \left [-{\nu_{1} M\over 1 - \tau }
Re ( y_{1} T_{2} + y_{2} T_{4} ) + 4 M Re
( z_{1} T_{2} + z_{2} T_{4} + z_{3} T_{6} ) \right ] \Biggr \},
\label {5.34}
\eeq
$$
Y_{pp}=-\; \Biggl \{ (\alpha_{1}^2 \alpha_{3} +
\nu_{3}) [(1-\tau) |T_{1}|^2+2 \nu_{1} M Re(T_{1}
T_{2}^{\ast}) + M^2 (\nu_{1}^2 - \nu_{2}^2) |T_{2}|^2]
$$
\beq
+\; (\alpha_{2}+\nu_{3}) [(1-\tau) |T_{3}|^2+2 \nu_{1} M
Re(T_{3} T_{4}^{\ast})+M^2(\nu_{1}^2-\nu_{2}^2) |T_{4}|^2]
\label {5.35}
\eeq
$$
~ + \; ( \alpha_{1}^2 \alpha_{3} + \alpha_{2} + 2 \nu_{3} ) \tau \left (
 - {|T_{5}|^2 \over M^4 \nu_{2}^2} + { M^2 \over \alpha_{3} }
|T_{6}|^2 \right ) \Biggr \} \; {16 M^4 \over r^{4}} \; .
$$

For the invariant variables in Eqs. (\ref{5.29}) -- (\ref{5.35}) used in
determining the Bethe-Heitler term ($Y_{ee}$), the interference term
($Y_{ep}$), and the term corresponding to proton brems\-strahlung ($Y_{pp}$),
we used the notation adopted in \cite{b64}:
\beqn
&&y_{1} = 2 \alpha_{1} [ \alpha_{1} \alpha_{3} ( \nu_{2} \nu_{5} - \nu_{1}
 \nu_{4} ) + 2 \nu_{4}^2 + \nu_{2} \nu_{3} ] , \; \nu_{1} = k q_{+}/2M^{2}\; ,
 \; \nu_{2} = - kq_{-}/ 2 M^2 \; ,\nonumber \\
&&y_{2} = 2 \alpha_{2} \; ( \nu_{2} \nu_{5} - \nu_{1} \nu_{4} ) - \alpha_{1}
 \nu_{2}^2  , \; \nu_{3} = r^2 / 4 M^2  , \; \nu_{4} = kq_{+}/ 4 M^2 \; ,
\; \nu_{5} = p_{+} q_{+} / 4 M^2  ,\nonumber \\
&&y_{3}=-(4 \nu_{3} / \nu_{2}^2 ) \; [\alpha_{1} \alpha_{3} (\nu_{1} \nu_{2}
 ( \nu_{2} + \nu_{3} ) - 2 \nu_{4} ( \nu_{1} \nu_{4} - \nu_{2} \nu_{5} ) ) +
\nu_{4} ( 4 \nu_{4}^2 - \nu_{2}^2 ) ] \; , \nonumber \\
&&\alpha_{1}=\nu_{5}+\nu_{1}\nu_{4}(2 \nu_{3}+\nu_{2}) / \nu_{2}^2 \; ,
\; \alpha_{3}=\nu_{2}^2 / (\nu_{2}^2+(\nu_{2}+\nu_{3})(\nu_{1}^2 -
\nu_{2}^2 ) ) \; , \nonumber \\
&&\alpha_{2} = m^2/M^2-\nu_{3}+M^6/D^2[-(\nu_{1} \nu_{4}+\nu_{2}
\nu_{5})^2+4\nu_{3}(\nu_{4}^2-\nu_{1} \nu_{4} \nu_{5})-4\nu_{3}
\nu_{4}^2(\nu_{2}+\nu_{3})] \; ,\nonumber \\
&&D^2 = M^6 \; ( \nu_{2}^2 + ( \nu_{2} + \nu_{3} ) ( \nu_{1}^2 - \nu_{2}^2 ) ) =
M^6 \nu_{2}^2 / \alpha_{3} \; , \; \lambda_{1} = p_{1} k \; , \; \lambda_{2}
= p_{2} k \; ,\nonumber \\
&&z_{1} = \nu_{1} \nu_{4} \alpha_{1}^2 \alpha_{3} \; ,
 \; z_{2} = \nu_{2} \nu_{4} \alpha_{2} \; , \;
z_{3} = 1/4 \alpha_{1} \; ( 2 \nu_{2} ( 2 \alpha_{2} + \nu_{2} + \nu_{3} ) +
 4 \nu_{4}^2 - \nu_{2}^2 ) \; .\nonumber
\eeqn

It should be noted that the expression obtained for the
differential cross section (\ref{5.29}) coincides, within
the definition of the initial quantities (the tensor $M_{\mu \nu}$),
with the result obtained in \cite{b64}, if one expresses
in the latter the form factors $f_{1}$ and $f_{2}$ through
$g_{e}$ and $g_{m}$.  Nevertheless, the Bethe-Heitler term
$Y_{ee}$ and the interference term $Y_{ep}$ have a more compact form
due to the factorization of the electric and magnetic form factors.

Let us consider contributions of all three graphs to the cross section
for the reaction (\ref{5.1}) in the selected kinematics when the
initial proton is at rest ($q_{1} = (M,0)$), and the electron beam
energy is $E_{e} = 200$ \MEV. Performing the required integration
over the phase space we obtain \cite{b65}:
\beq
d \sigma = { \alpha^{3} \omega ^2 \mid \vec q_{2} \mid T \mid ^{2}
 \over 16 \pi^{2} M \mid \vec p_{1} \mid (p_{2} k) } \;
\;  d E_{pk} \; d \Omega_{q_2} \; d \Omega_{\gamma} \; ,
\label {5.36}
\eeq
where $d \Omega_{\gamma}$ and $d \Omega_{q_2}$ are the elements of the photon
and proton solid angles, and $E_{pk}$ is the kinetic energy of the recoil
proton. The differential cross section (\ref{5.36}) was calculated
numerically in the region $5 \leq E_{pk} \leq 35$ \MEV with the sum
and the difference of the electric ($\alpha_{p}$) and magnetic
($\beta_{p}$) polarizabilities equal to $\alpha_{p}+ \beta_{p} = 14$
and $\alpha_{p} - \beta_{p} = 10$ (in units of $10^{-4}fm^3$)
\cite{b58}-\cite{b60}. We assume that the reaction kinematics is
planar, and that the photon emission and proton scattering angles are
$\vartheta_{\gamma}=135^{0}$ and $\vartheta_{p}=- 20.5^{0}$,
respectively (all angles are measured from the direction of the
primary electron beam). Calculations show that in the
entire range of proton kinetic energy considered, $5 \leq E_{pk} \leq
35$ \MEV, for the selected angles $\vartheta_{\gamma} =135^{0}$ and
$\vartheta_{p} = - 20.5^{0}$, the electron scattering angle
$\vartheta_{e}$ and the 4-momentum transfer $ |r| =
\sqrt{-(p_{2}-p_{1})^2}$ are bounded by the values $| \vartheta_{e}|
\leq 6.4^{0}$ and $|r| \leq 7.3$ \MEV, with the minimum value of
$|r|$ corresponding to forward electron scattering.

\begin{figure}[ht]
\vspace{2.8cm}
\centerline{\epsfxsize=0.60\textwidth\epsfbox[10 10 550 600]{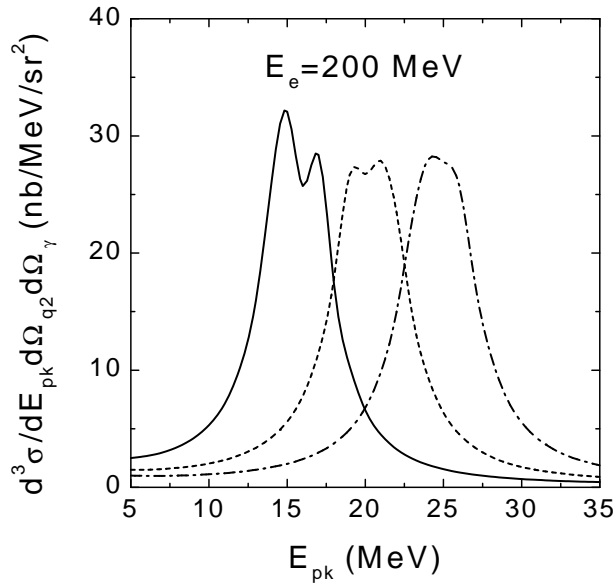}}
\vspace{-6.4cm}
\caption
{Differential cross section (\ref{5.36}) for the reaction (1)
in the kinematics where proton bremsstrahlung dominates (see comments
in the text). Proton scattering  angles are $\vartheta_p=-20.5^0$
(solid line), $\vartheta_p= -20.0^0$ (dashed line),
$\vartheta_p=-19.5^0$ (dot-dashed line) and photon emission
angle is $\vartheta_{\gamma}= 135^0$.}
\end{figure}

The results of numerical calculations of the cross section
(\ref{5.36}), $d \sigma / d E_{pk} / d \Omega_{q_2} / d \Omega_{\gamma}$
in the above kinematics are shown in Fig.2.  We see that in the angular
range studied the cross section
for the reaction $ep \to ep \gamma$ has a sharp peak consisting of
two maxima. This peak originates from the factor $1/r^4$ in Eq.
(\ref{5.35}) for $Y_{pp}$. The two maxima have a kinematical origin
and arise from the interference of two pole graphs corresponding to
quasi-real Compton scattering. The cross section (\ref{5.36}) has a
strong angular dependence, which, in particular, causes the two
maxima to disappear when the proton scattered (or photon emission)
angle is changed by only one degree (i.e., for $\vartheta_{p} = -
19.5^{0}$), so that we have an ordinary peak at $E_{pk} = 25$ \MEV.

The differential cross section (\ref{5.36}) shown in Fig.2,
is the sum of the Bethe-Heitler ($\sigma_{ee}$), the interference
($\sigma_{ep}$), and the proton ($\sigma_{pp}$) terms (see
(\ref{5.30})), where the symbol ($\sigma$) denotes the cross section
of the form (\ref{5.36}) with $\mid T \mid^2$ replaced by $Y_{ee}$,
$Y_{ep}$, and $Y_{pp}$, respectively. Numerical calculations show
that in the entire range of proton kinetic energy studied, $5 \leq
E_{pk} \leq 35$ \MEV, the ratios of the Bethe-Heitler term
$\sigma_{ee}$ and the interference term $\sigma_{ep}$ to the term
corresponding to proton emission $\sigma_{pp}$ are bounded by the
values $\sigma_{ee} /\sigma_{pp}< 0.02$ and $| \sigma_{ep} | /
\sigma_{pp} < 0.05$. The calculations carried out for another set of angles
($\vartheta_{\gamma} =135^{0}$ and $\vartheta_{p} = - 20^{0}$) give results
which are only slightly different: $\sigma_{ee} / \sigma_{pp}< 0.05$
and $|\sigma_{ep} | / \sigma_{pp} < 0.075$. Since these ratios are much
smaller than unity, the main requirement (see \cite{b60}) for separation
of the background, which is mainly electron bremsstrahlung, is satisfied.

To investigate the sensitivity of the reaction (1)
to the proton polarizability we performed numerical
calculations of the cross section (\ref{5.36}) for the same set of
angles ($\vartheta_{\gamma} = 135^{0}$ and $\vartheta_{p} = -
20^{0}$) and  fixed sum of the electric and magnetic
polarizabilities $\alpha_{p} + \beta_{p} = 14$ but different values
of the difference:  (a) $\alpha_{p} - \beta_{p} = 10$ and (b)
$\alpha_{p} - \beta_{p} = 6$.  It turned out that the cross section
(\ref{5.36}) is about 8\% larger for the smaller difference of
polarizabilities. Therefore, in this kinematics the cross section for
the reaction $ e p \to e p \gamma$ is quite sensitive to the proton
polarizability \cite{b65}.

\section{\bf Emission of a linearly polarized photon by an electron
in the reaction $ep \to ep \gamma$}

Let us consider now the emission of a linearly polarized photon by an
electron in the reaction $ ep \to ep \gamma$, taking into account the
proton recoil and form factors. Our study will be limited to the
contribution of the two Bethe-Heitler graphs (a) and (b) in Fig.1,
which corresponds to the matrix element (\ref{5.2}). The contribution
of the graph with VCS on a proton can be neglegted when the initial
electrons have ultrarelativistic energies, and the photon and final
electron are scattered at small forward angles ($\vartheta_{\gamma}
\sim m/E_{e}, \; \vartheta_{e} \sim m/E_{e}, \; m/E_{e} \ll 1 $).

We are interested in these effects for the following reasons. First, even
though the Bethe-Heitler process has been intensively studied earlier
in the case of the emission of linearly polarized photons
\cite{b69,b70} and is widely used to obtain them at accelerators
\cite{b71}, up to now the proton recoil and form factors have not
been accurately taken into account (in contrast to the unpolarized
case). Second, as was shown in \cite{b72}, the inclusion of these
factors in the case of unpolarized photons leads to a strong change
of the differential cross section for the Bethe-Heitler process.
Since the polarization characteristic of the scattered radiation are
expressed in terms of the differential cross section for the emission
of an unpolarized photon (see below), it is clear that inclusion of
the recoil and form factors is essential.

The covariant expression for the differential cross section for the
Bethe-Heitler process (in the Born approximation) taking into account the
proton recoil and form factors in the case of emission of a linearly polarized
photon has been obtained by us in \cite{b73}. It has the form
\beq
d \sigma_{BH} = { \alpha^{3} \mid T_{e} \mid ^{2} \delta^{4} ( p_{1} + q_{1} - p_{2}
- q_{2} - k ) \over 2 \pi^{2} \sqrt{ (p_{1} q_{1})^{2} - m^{2} M^{2}}} \;
\; { d^{3} \vec p_{2} \over 2 p_{20}} \; {d^{3} \vec q_{2} \over 2 q_{20}}
{d^{3} \vec k \over 2 \omega} \; \; ,
\label {6.1}
\eeq
\beq
\mid T_{e} \mid^2 = {4 M^{2}\over q^{4}} \; ( \; g_{e}^{2} \; Y_{I}^{~e}
 + \tau \; g_{m}^{2} \; Y_{II}^{~e} \; ) \; ,
\label {6.2}
\eeq
\beq
Y_{I}^{~e} = 2 - {\lambda_{1} \over \lambda_{2}} - {\lambda_{2} \over \lambda_{1}}
- {\tau \over 1 - \tau} \; {(k q_{+})^2 \over \lambda_{1} \lambda_{2}} +
 q^2 \; (e a)^2 + 4 \; (e A)^2 \; ,
\label {6.3}
\eeq
\beq
Y_{II}^{~e} = - 2 - {\lambda_{1} \over \lambda_{2}} - {\lambda_{2} \over \lambda_{1}}
+ {\tau \over 1 - \tau} \; {(k q_{+})^2 \over \lambda_{1} \lambda_{2}} +
 ( q^2 + 4 m^2 ) \; (e a)^2 - 4 \; (e A)^2 \; ,
\label {6.4}
\eeq
\beq
a = {p_{1} \over \lambda_{1}} - {p_{2} \over \lambda_{2}} \; \; ,
\; \; A = b_{0} + {(b_{0} p_{2}) p_{1} \over \lambda_{1}} - {(b_{0} p_{1}) p_{2}
 \over \lambda_{2}} \; .
\label {6.5}
\eeq
All the quantities entering (\ref{6.1})-(\ref{6.5}) are defined in
the previous section. Thus, the differential cross section for
the Bethe-Heitler process in the case of emission of a linearly
polarized photon $d \sigma_{BH}$ (\ref{6.1}) is naturally splitted
into the sum of two terms containing only the squares of the Sachs
form factors and corresponding to the contribution of transitions
without ($\sim g_{e} ^2 \; Y_{I}^{~e}$) and with ($\sim \tau \;
g_{m}^2 Y_{II}^{~e}$) proton spin flip.

Let us discuss the properties of the 4-vector $a$, which is well known from
the theory of emission of long-wavelength photons \cite{b10}, and the 4-vector
$A$. They both satisfy a condition which follows naturally from the
requirement of gauge invariance: $a\; k = A \; k = 0$, and, in addition,
they are spacelike vectors: $a^2<0 $ and $A^2<0$. This is easily verifed
by using the 4-momentum conservation law and the explicit form of $a^2$
and $ A^2$:
$$
a^2 = m^2 \; \left ( {1 \over \lambda_{1}} - {1 \over \lambda_{2}} \right )^2
+ {r^2 \over \lambda_{1} \lambda_{2}} \; ,
\label {6.6}
$$
$$
A^2 = 1 + {m^2 \over 4 M^2 (1-\tau)}\; \left ({q_{+}p_{1} \over \lambda_{2}}
- {q_{+}p_{2} \over \lambda_{1}} \right ) + {\tau \over 1-\tau} \; {q_{+}p_{1}
\cdot q_{+}p_{2} \over \lambda_{1} \lambda_{2}} \; .
\label {6.7}
$$
We note that the 4-vector $A$ was first introduced in \cite{b73}.

Using the electron 4-momenta $p_{1}$ and $p_{2}$ and the photon 4-momenta $k$,
we construct the 4-vectors of the photon linear polarization $e_{\parallel}$
and $e_{\perp} \;(e_{\parallel} k=e_{\perp}k=e_{\parallel}e_{\perp}=0$):
$$
e_{\parallel} = {(p_{2} k) p_{1} - (p_{1} k ) p_{2} \over \rho^{'} } \; , \; e_{\perp}
 = {[p_{1} \cdot p_{2}]^{\times} k \over \rho^{'}} \; ,
\label {6.8}
$$
where $\rho^{'}$ is determined from the normalization conditions:
$e_{\parallel}^2 = e_{\perp}^2 = - 1$. Then the degree of photon linear
polarization will be given by the following expressions \cite{b73}:
\beq
P_{\gamma} = { \mid T_{\perp} \mid^2 - \mid T_{\parallel} \mid^2 \over
\mid T_{\perp} \mid^2 + \mid T_{\parallel} \mid^2 } = {A_{1} \over A_{2}} \; ,
\label {6.9}
\eeq
where
\beq
A_{1} = {16 \;M^2 \over q^4} \; ( g_{e}^2 \; A_{11} + \tau \; g_{m}^2 \;
 A_{12} ) \; ,
\label {6.10}
\eeq
\beq
A_{2} = {8 \;M^2 \over q^4} \; ( g_{e}^2 \; Y_{1} + \tau \; g_{m}^2 \;
 Y_{2} ) \; ,
\label {6.11}
\eeq
\beq
Y_{1} = 2 - {\lambda_{1} \over \lambda_{2}} - {\lambda_{2} \over \lambda_{1}}
- {\tau \over 1 - \tau} \; {(k q_{+})^2 \over \lambda_{1} \lambda_{2}} -
 2 \; \tau M^2 \; a^2 - 2 \; A^2 \; ,
\label {6.16}
\eeq
\beq
Y_{2} = - 2 - {\lambda_{1} \over \lambda_{2}} - {\lambda_{2} \over \lambda_{1}}
+ {\tau \over 1 - \tau} \; {(k q_{+})^2 \over \lambda_{1} \lambda_{2}} -
 2 \; \tau \; M^2 \; a^2 + 2 \; A^2 - 2 m^2 \; a^2 \; .
\label {6.17}
\eeq
$$
A_{11} = A^2 + \tau \; M^2 \; a^2 + 2 (e_{\perp} b_{0})^2 \; ,
\nonumber
$$
$$
A_{12} = - A^2 + \tau \; M^2 \; a^2 - 2 (e_{\perp} b_{0})^2 + m^2 \; a^2 \; ,
$$
$$
(e_{\perp} b_{0})^2 = - \; {4 (SD)^2 \over M^2 (1-\tau) a^2 \lambda_{1}^2
 \lambda_{2}^2 } \; ,
$$
$$
SD = 1/2 \; \epsilon_{\mu \nu \rho \sigma} (p_{1})^{\mu} (p_{2})^{\nu} (q_{1})
^{\rho} (q_{2})^{\sigma} \; ,
$$
It is easy to check that $A_{2}$ (\ref{6.11}) coincides with the expression
for $Y_{ee}$ (\ref{5.31}) determining the Bethe-Heitler cross section
in the case of unpolarized particles: $ A_{2} = Y_{ee}$, and also that
$Y_{1} = Y_{I}$ and $Y_{2} = Y_{II}$ (see (\ref{5.32}) and
(\ref{5.33})).

Therefore, owing to the factorization of the squared form factors
$g_{e}$ and $g_{m}$ and also the use of the 4-vectors $a$ and $A$
(\ref{6.5}), the differential cross section for the Bethe-Heitler process
both for linearly polarized photon (\ref{6.2}) and
unpolarized photon (\ref{6.11}), (\ref{5.31}), can be written in a
rather compact form.

An integration of Eq. (\ref{6.1}) over $d^3 \vec q_2$ and $d
p_{20}$ in the rest frame of the initial proton ($q_1 =(M,0)$) gives
the following result:
\beq
{ d \sigma_{BH} \over d \omega \; d \Omega_{\gamma} \; d \Omega_{e} } =
{\alpha^3 \; \omega \over (2\pi)^2 } \; {|\vec p_2| \over |\vec p_1| } \;
{| T |^2 \over q^4 \; } \; ,
\label {6.18}
\eeq
\beq
| T |^2 =  g_{e}^{2} \; Y_{I}^{~e} + \tau \; g_{m}^{2} \; Y_{II}^{~e} \; .
\label {6.19}
\eeq
Let us consider the limit of the cross section (\ref{6.18}) when the proton
is a pointlike (structureless) particle with infinite mass, i.e., we
assume that $g_{e} = g_{m} = 1$ and $q_2 = (M, \vec q) \simeq (M,0)$,
where $\vec q = \vec p_1 - \vec p_2 - \vec k$ is the momentum
transferred to the proton.  In this limit ($M \to \infty$), $E_{kp} =
\vec q~^2/2M \to 0, \; \vec q /2M \to 0$, and $b_0 = (1, \vec q/2M)
\simeq (1,0)$. We choose the Coulomb gauge for the photon
polarization vectors: $e = (0, \vec e)$ in which one obtains $$ eb_0
= 0 , \; ea = {p_1 e \over \lambda_1 } - {p_2 e \over \lambda_2 }, \;
eA = p_{20}\; {p_1 e \over \lambda_1 } - p_{10} \; {p_2 e \over
\lambda_2 }, \; \tau (q_{+} k)^2 = \omega^2 q^2\;.
$$
Using these expressions we have in the above limit for
(\ref{6.19}):
\beq
|T|^2 = 2 - {\lambda_{1} \over \lambda_{2}} - {\lambda_{2} \over \lambda_{1}}
- {\omega^2 q^2 \over \lambda_{1} \lambda_{2}} +
 q^2 \; (e a)^2 + 4 \; (e A)^2 \; ,
\label {6.20}
\eeq
or, in expanded form,
$$
|T|^2 = 2 - {\lambda_{1} \over \lambda_{2}} - {\lambda_{2} \over \lambda_{1}}
- {\omega^2 q^2 \over \lambda_{1} \lambda_{2}} +
\; (\;4 p_{20}^2 + q^2\; ) \left ({p_1 e \over \lambda_1 }\right )^2
$$
\beq
+ \; (\; 4 p_{10}^2 + q^2\;) \left ({p_2 e \over \lambda_2 }\right )^2 - 2 \; (\; 4 p_{10} p_{20} + q^2\; ) \;
{p_1e \cdot p_2e \over \lambda_{1} \lambda_{2} }\; .
\label {6.21}
\eeq
The expressions (\ref{6.18}), (\ref{6.21}) for the differential
cross section for the Bethe-Heitler process $d \sigma_{BH} / d \omega / d
\Omega_{\gamma} / d \Omega_{e}$ in the limit where the proton is an
infinitely heavy, structureless particle coincide with the result of
\cite{b69}.

\section{\bf Virtual-photon polarization in the reaction\\
$ep \to ep \gamma \; (ep \to eX) $ }

The reactions $ep \to ep \gamma$ and VCS on the proton have recently
become interesting not only at low and intermediate energies
\cite{b60}, but also at high electron energies and 4-momenta
transferred to the proton \cite{b63}, \cite{b74}-\cite{b77}. The VCS
offers greater possibilities for studying hadronic structure
than the RCS process, because in it the energy and three-momentum
transferred to the target can be varied independently. These
attractive properties of VCS have led to the suggestion that it could
be used for experimental study of the nucleon structure
\cite{b74,b75} and have made it necessary to perform a thorough
theoretical study of the reaction $ep \to ep \gamma$ (see
\cite{b63,b76,b77} and references therein). To calculate VCS on the
proton, it is necessary to know the hadron ($W_{\mu \nu}$) and lepton
($L_{\mu \nu}$) tensors \cite{b63,b78}:  \beq L_{\mu \nu}=J_\mu
J_\nu^*,~~J_\mu=\overline{u}(p_2)\gamma_\mu u(p_1) \; , \label {7.1}
\eeq
where $u(p_{i})$ are electron bispinors, $\overline{u}(p_{i}) u(p_{i}) =
2 m$, and $m$ is the electron mass ($i =1,2$). The interpretation of the
results is considerably simplified if the tensor $L_{\mu \nu}$ is expressed
in terms of the longitudinal and transverse polarization vectors of the
virtual photon. The corresponding expressions can be found in \cite{b63} and
\cite{b78}. However, they have two defects: (1) the electron mass
is neglegted, which is of course justified at ultrarelativistic electron
energies and large squared 4-momentum of the virtual photon; (2) they have
a noncovariant form. A lepton tensor free of these defects was constructed
in \cite{b79a}.

Let us consider the question of the polarization state of a virtual
photon with 4-momentum $r=p_{1}-p_{2}$ which is exchanged between the
electron and proton in the reaction $ep \to ep \gamma$ (see Fig.1c). Using
the vectors of the orthonormal basis $a_{A}$ (\ref{5.9}) ($A=(0,1,2,3)$):
\beq
a_{0} = p_{+}/\sqrt{p_{+}^{2}} \; , \; a_{3} = p_{-}/\sqrt{-p_{-}^{2}} \; ,
\; a_{2} = [a_{0} \cdot a_{3}]^{\times} q_{1} /\rho \; , \;  a_{1} = [a_{0}
\cdot a_{3}]^{\times} a_{2} \; ,
\label {7.2}
\eeq
$$
p_{\pm} = p_{2} \pm p_{1} \; ,
 \; a_{2} q_{1} = 0 \; , \; a_{1}^{2} =
a_{2}^{2} = a_{3}^{2} = - a_{0}^{2} = - 1 \; ,
$$
which satisfies the completeness relation
\beq
a_{0} \cdot a_{0} - a_{1} \cdot a_{1} - a_{2} \cdot a_{2} - a_{3} \cdot a_{3}
= g \; ,
\label {7.3}
\eeq
where $g=(g_{\mu\nu})$ is the metric tensor with signature $g_{\mu\nu}=
(+---)$, we construct the 4-vectors of the longitudinal ($e_{3}$) and
transverse ($e_{1}, \; e_{2}$) polarization of a virtual photon with
4-momentum $r$ \cite{b79a}:
\beq
e_{1} = {[a_{0} \cdot a_{1}] q_{1} \over \sqrt{(a_{3}q_{1})^2 + q_{1}^2} }
\; , \; e_{2} = a_{2} = { [a_{0} \cdot a_{3}]^{\times}q_{1} \over \rho} \;
, \; e_{3} = { (1 + a_{3} \cdot a_{3})q_{1} \over \sqrt{(a_{3}q_{1})^2
+ q_{1}^2} } \; ,
\label {7.5}
\eeq
where
$$
\rho^2 = (a_{1}q_{1})^2= { 2p_{1}p_{2} \cdot p_{1}q_{1} \cdot p_{2}q_{1}
- M^2 ((p_{1}p_{2})^2 - m^4) - m^2 ((p_{1}q_{1})^2 + (p_{2}q_{1})^2) \over
 (p_{1}p_{2})^2 - m^4} .
$$
It is easily verifed that the 4-vectors $e_{i} \; (i=1,2,3)$ are orthogonal
to each other ($e_{i} e_{j} = 0, \; i \neq j$), and also that $e_{i}r
= e_{i}a_{3}= 0$ and $e_{1}^2 = e_{2}^2 = - e_{3}^2 = -1$. The 4-vectors
$e_{i}$ (\ref{7.5}) are not changed when the auxiliary 4-vector $q_{1}$
is replaced by $q_{1} + p_{1} - p_{2} = q_{2} + k$ (since $p_{1} - p_{2} =
r =-2y a_{3}$, where $y= \sqrt{-r^2}/2$, and the vectors
$a_{A}$ (\ref{7.2}) are orthogonal). For this reason, the
virtual-photon polarization vectors $e_{i}$ (\ref{7.5}) in the rest
frame of the incident proton or in the c.m. frame of the final proton
and photon can be considered as equivalent and their use lead to the
same expressions.  Below we restrict ourselves to the rest
frame of the incident proton, $q_{1}= (M,0,0,0)$, where the 4-vectors
$e_{i}$ have the form:
\beq
e_{1} = (0,1,0,0), \; e_{2} = (0,0,1,0), \; e_{3} = {1 \over \sqrt{-r^2}}
(\mid \vec r \mid, r_{0} \vec n_{3}) \; ,
\label {7.6}
\eeq
where $\vec n_{3}$ is a unit vector directed along $\vec r \;(\vec
n_{3}^{~2} =1)$, and $r_{0}$ is the time component of the 4-vector
$r=(r_{0}, \vec r)$.

The four mutually orthogonal vectors $e_{1}, e_{2}, e_{3}$, and $a_{3}$
also satisfy the completeness relation:
\beq
e_{3} \cdot e_{3} - e_{1} \cdot e_{1} - e_{2} \cdot e_{2} - a_{3} \cdot a_{3}
= g \; ,
\label {7.7}
\eeq
which allows $a_{0}$ and $a_{1}$ to be expressed in terms of $e_{1}$ and
$e_{3}$:
\beq
a_{1}=\alpha e_{3} - \beta e_{1} \; , \; a_{0} = \beta e_{3} - \alpha e_{1}\; ,
\; \beta^2=1 + \alpha^2 \; \; ,
\label {7.8}
\eeq
\beq
\alpha = e_{3} a_{1} = a_{0} e_{1} = { a_{1}q_{1} \over \sqrt{(a_{3}q_{1})^2
 + q_{1}^2} }\; , \; \beta = e_{1} a_{1} = e_{3} a_{0} = {a_{0} q_{1} \over
\sqrt{(a_{3}q_{1})^2 + q_{1}^2} } \; .
\label {7.9}
\eeq
In the DSB (\ref{4}) the matrix elements of the electron current have the form
of (\ref{5.18}). Let us write them in terms of the 4-vectors $e_{i}$
(\ref{7.5}) \cite{b79a}:
\beq
( J^{\delta,\delta}_{e} )_{\mu}= 2 m \;  (\beta e_{3} - \alpha e_{1})_{\mu} \; , \;
( J^{-\delta,\delta}_{e} )_{\mu}  = - 2 \delta y \; (\alpha e_{3} - \beta e_{1} + i \delta e_{2})_{\mu} \; .
\label {7.11}
\eeq
Therefore, for spin-non-flip transitions $(J^{\delta,\delta}_{e})$ the
virtual-photon polarization vector is a superposition of the longitudinal
($\beta e_{3}$) and transverse linear ($-\alpha e_{1}$) polarizations,
while for spin-flip transitions $(J^{-\delta,\delta}_{e})$ it is a
superposition of the longitudinal ($\alpha e_{3}$) and transverse
elliptical [$e_{\delta} =(0, \vec e_{\delta})= -\beta e_{1} + i
\delta e_{2}$] polarizations. Here the state of a photon with
elliptical polarization vector $e_{\delta} =(0, \vec e_{\delta})$ has
degree of linear polarization (equal to the ratio of the difference
and sum of the squared semiaxes) \cite{b79a}:

\beq
\kappa_{\gamma} = { \beta^2 - 1 \over \beta^2 + 1} = { \alpha^2 \over \beta^2
 + 1} \; .
\label {7.12}
\eeq
Inverting this relation, we obtain:
$$
\beta^2 = { 1 + \kappa_{\gamma} \over 1 - \kappa_{\gamma}} \; \; , \;
\alpha^2= {2 \kappa_{\gamma} \over 1 - \kappa_{\gamma}} \; \; .
\label {7.13}
$$
Now we find the squared moduli of the vectors $\vec e_{\delta}$ and
$\vec a_{\delta}$:
$$
\mid \vec e_{\delta} \mid^2 = 1 + \beta^2 = {2 \over 1 - \kappa_{\gamma}} \; ,
\; \mid \vec a_{\delta} \mid^2 = ( 1 + \beta^2 ) \; (1 + \kappa_{L}) \; ,
\label {7.14}
$$
\beq
\kappa_{L} = \kappa_{\gamma} \vec e_{3}^{~2} = \kappa_{\gamma} { r_{0}^2 \over
(-r^2)} \; , \; \vec e_{3}^{~2} = { r_{0}^2 \over (-r^2)} \; .
\label {7.15}
\eeq
Let us introduce the normalized vectors $\vec e_{\delta}~'$ and
$\vec a_{ \delta}~'$:
\beq
\vec e_{\delta}~' = {\vec e_{\delta} \over \sqrt{1 + \beta^2}} =
\sqrt {{1 - \kappa_{\gamma} \over 2}} \; \vec e_{\delta} \; , \;
|\vec e_{\delta}^{~'}|^2 = 1 \; .
\label {7.16}
\eeq
\beq
\vec a_{\delta}~' = {\vec a_{\delta} \over \sqrt{1 + \beta^2}} =
\sqrt {{1-\kappa_{\gamma} \over 2}} \; \vec a_{\delta} \; , \;
|\vec a_{\delta}^{~'} |^2 = 1 + \kappa_{\gamma} \vec e_{3}^{~2} = 1 +
\kappa_{L} \; ,
\label {7.17}
\eeq
It is seen that the elliptical-polarization vector $\vec
e_{\delta}$ of a virtual photon can be normalized to unity ($|\vec
e_{\delta}~'|^2 = 1$), but the presence of a longitudinal
polarization makes this normalization impossible for the total vector
$\vec a_{\delta}~'$ simultaneously. The quantity $\kappa_{L}$
(\ref{7.15}) corresponding to the inequality $|\vec a_{\delta}~' |^2
= 1 + \kappa_{L} \neq 1$ has the meaning of the degree of
longitudinal polarization of a virtual photon emitted in a transition
with electron spin flip. In the ultrarelativistic limit, when the
electron mass can be neglected, the quantities $\kappa_{\gamma}$ and
$\kappa_{L}$ can be interpreted as the total degrees of linear and
longitudinal polarization of the virtual photon. In this (massless)
case we have:  \beq (a_{3}q_{1})^2 + q_{1}^2 = - M^2\; { \vec r^{~2}
\over r^2} \; , \; (a_{1} q_{1}) ^2 = M^2 \; ctg^2 \vartheta /2 \; ,
\label {7.18}
\eeq
\beq
\kappa_{\gamma}^{-1} = 1 - 2 \; {\vec r^{~2} \over r^2} \; tg^2
\vartheta /2 \; ,
\label {7.19}
\eeq
where $\vartheta$ is the angle between the vectors $\vec p_{1}$ and
$\vec p_{2}$. Equation (\ref{7.19}) for $\kappa_{\gamma}$ coincides
with the result of \cite{b78}.

The vector $\vec a_{\delta}~'$ (\ref{7.17}) can also be written as
$$
\vec a_{\delta}~' = \sqrt{\kappa_{L}} \; \vec n_{3} - \sqrt {{1 +
\kappa_{\gamma} \over 2}} \; \vec e_{1} + i \delta \; \sqrt {{1 -
\kappa_{\gamma} \over 2}} \; \vec e_{2} \; \; ,
$$
which makes it easy to construct the polarization density matrix for a virtual
photon in the massless limit (both in the polarized case, which for massless
particles is helical polarization, and in the unpolarized case; see
\cite{b78}).

To obtain the complete expression for $\kappa_{\gamma}$ and $\kappa_{L}$
arising from the contributions of the matrix elements both without and with
spin flip, we construct the lepton tensor averaged over electron spin states.
Using the matrix elements (\ref{5.18}) this can be easily
done \cite{b79a}:
\beq
\overline{L}_{\mu \nu}= 4 m^2 \; (a_{0})_{\mu} (a_{0})_{\nu} + 4 y^2 \;
( (a_{1})_{\mu} (a_{1})_{\nu} + (a_{2})_{\mu} (a_{2})_{\nu} ) \; .
\label {7.21}
\eeq
Using the completeness condition (\ref{7.3}) and gauge invariance, the tensor
$\overline{L}_{\mu \nu}$ can be written as
\beq
\overline{L}_{\mu \nu} = 4 x^2 \; (a_{0})_{\mu} (a_{0})_{\nu} - 4 y^2 \;
g_{\mu\nu} \; ,
\label {7.22}
\eeq
where $x^2= m^2 + y^2$. The tensor $\overline{L}_{\mu \nu}$ (\ref{7.22})
is used to reduce the calculation of the contribution of graphs with
VCS on a proton to the cross section for the reaction $ep \to ep \gamma$
to calculation of the trace of a product of tensors:
\beq
Y_{pp} = \overline{L}_{\mu \nu} \; W_{\mu \nu} \; , \; W_{\mu \nu} = V_{\mu}
V_{\nu}^{\ast} \; ,\; V_{\mu} = \overline{u}(q_{2}) \; M_{\mu \nu}
 e^{\nu} \; u(q_{1}) \; {1\over r^2} \; .
\label {7.23}
\eeq
Let us express the tensor $\overline{L}_{\mu \nu}$ (\ref{7.21}) in the
terms of the virtual-photon polarization vectors $e_{i}$ (\ref{7.5}).
As a result, it naturally breaks up into the sum of three terms corresponding
to the contributions of transverse ($L_{T}$) and longitudinal ($L_{L}$)
states and their interference ($L_{LT}$) \cite{b79a}:
\beqn
&&\overline{L} = 4y^2 \; ( L_{T} \; + \; L_{L} \; + \; L_{LT} \; ) \; ,
\label {7.24}\\
&&L_{T} = e_{1} \cdot e_{1} \; (\beta^2 +  \alpha^2  m^2 / y^2 ) +
                 e_{2} \cdot e_{2}  \; ,
\label {7.25}\\
&&L_{L} = e_{3} \cdot e_{3} \; ( \alpha^2 + \beta^2  m^2 / y^2 ) \; ,
\label {7.26}\\
&&L_{LT}= - \; ( e_{1} \cdot e_{3} + e_{3} \cdot e_{1} ) \;\alpha \beta \;
 (1 + m^2/y^2) \; .
\label {7.27}
\eeqn
Then the total degree of linear polarization of the virtual photon is
given by
\beq
\kappa_{\gamma} ' = { \beta^2 + \alpha^2 m^2/y^2 - 1 \over \beta^2 +
\alpha^2 m^2/y^2 +1} = { \alpha^2 \over \beta^2 +1 - 2 m^2/x^2} \; .
\label {7.28}
\eeq
Since $\alpha$ and $\beta$ are the same in Eqs. (\ref{7.12}) and (\ref{7.28})
(see (\ref{7.9})), the inclusion of the electron mass in the
ultrarelativistic limit leads only to a slight increase of
$\kappa_{\gamma}$ \cite{b79a}:

\beq
\kappa_{\gamma} ' \simeq  \kappa_{\gamma} \left (1 + {2m^2 \over x^2 (1
+ \beta^2)}  \right ) \; .
\label {7.29}
\eeq
Inverting the relation in (\ref{7.28}), we find
\beq
\beta^2 + \alpha^2 m^2/y^2 = { 1 + \kappa_{\gamma} ' \over 1 -
\kappa_{\gamma} '} \; , \; \alpha^2 + \beta^2 m^2/y^2 = { 2 \kappa_{\gamma} '
 \over 1 - \kappa_{\gamma} '} + {m^2 \over y^2} \; .
\label {7.30}
\eeq
We can separate the completely polarized and unpolarized parts in the
transverse tensor:
$$
L_{T} = e_{1} \cdot e_{1} \, (\beta^2 +  \alpha^2  m^2 / y^2 - 1 ) +
e_{1} \cdot e_{1} +  e_{2} \cdot e_{2}\, =
 {2 \over 1 - \kappa_{\gamma}'}\, (\, \kappa_{\gamma}'
\, e_{1} \cdot e_{1} + (1 - \kappa_{\gamma}') \, ( e_{1} \cdot e_{1}
+  e_{2} \cdot e_{2} )/2 \,)\,.
\label {7.31}
\nonumber
$$
Therefore, the virtual-photon polarization density matrix $\rho_{ij}$ is
obtained from the tensor $\overline{L}_{ij}$ (\ref{7.24}) just as in the
massless case (see \cite{b78}):
\beq
\rho_{ij} =(1-\kappa_{\gamma}') \;  \overline{L}_{ij}/8 y^2 \;.
\label {7.32}
\eeq
For the degree of longitudinal polarization of the virtual photon we then
obtain:
\beq
\kappa_{L} ' = {r_{0}^2 \over (-r^2)} \kappa_{\gamma} ' \; \left ( 1 +
{m^2 \over y^2} \; { (1 - \kappa_{\gamma} ') \over 2 \kappa_{\gamma} '} \right ) \; .
\label {7.33}
\eeq
The expressions (\ref{7.28}) and (\ref{7.33}) for $\kappa_{\gamma} '$
and $\kappa_{L} '$ with $m = 0$ obviously become $\kappa_{\gamma}$ and
$\kappa_{L}$ of (\ref{7.12}) and (\ref{7.15}).

We conclude by noting that the region of applicability of the tensor
$\overline{L}_{\mu \nu}$ (\ref{7.24}) is not limited to only VCS on
the proton.  Since in fixed-target experimets the charged-lepton
scattering at available energies is mainly determined by virtual
photon exchange, the tensor $\overline{L}_{\mu \nu}$ (\ref{7.24}) can
also be used to study deep-inelastic electron scattering ($e^{\pm} p
\to e^{\pm} X$), and muon scattering ($\mu^{\pm}p \to \mu^{\pm} X$),
where inclusion of the mass is more important.

\section*{\bf Conclusion}

We have studied the reaction $ ep \to ep \gamma$
in the kinematics corresponding to electron scattering
at small angles and photon scattering at fairly large angles, where proton
bremsstrahlung dominates. The results of numerical calculations performed
in the rest frame of the initial proton at electron beam energy
$E_{e}=200$ MeV in the chosen kinematics show that the conditions
needed to separate the subprocess $\gamma p \to \gamma p$ from the
reaction $ep \to ep \gamma$ are satisfied, because the relative
contribution of the Bethe-Heitler and interference terms to the
reaction cross section is less than 10 \%, and the cross section for
the reaction $ep \to ep \gamma$ is quite sensitive to the proton
polarizability.

A compact expression was obtained for the differential cross section of the
Bethe-Heitler emission of a linearly polarized photon by an electron, taking
into account the proton recoil and form factors, owing to the factorization
of the squared electric and magnetic form factors of the proton. In the limit
where the proton is a pointlike particle of infinite mass, this
expression becomes to be the well-known one.

A covariant expression has been obtained for the lepton tensor in which the
contribution of states with transverse and longitudinal polarization of the
virtual photon is separated. It has been shown that inclusion of the lepton
mass tends to increase the degree of linear polarization of the virtual
photon.

\section*{\bf Acknowledgements}

The authors thank to A.I. L'vov for supplying them with
a computer code for numerical calculations of the proton RCS
amplitudes and for useful discussions. We are also indebted to V.A.
Petrun'kin for stimulating discussions of the results.

%\newpage

\end{document}